\documentclass[prl,twocolumn,superscriptaddress,showpacs]{revtex4}

\usepackage{graphicx}
\usepackage{dcolumn}
\usepackage{amsmath}
\usepackage{color}

\newcommand{\etal}{\textit{et al.}}

\begin{document}
\title{Coupling of localized moments and itinerant electrons in EuFe$_2$As$_2$ single crystals studied by Electron Spin Resonance}

\author{E.~Dengler}
\author{J.~Deisenhofer}
\email{joachim.deisenhofer@physik.uni-augsburg.de}
\author{H.-A.~Krug von Nidda}
\affiliation{Experimentalphysik V, Center for Electronic
Correlations and Magnetism, Institute for Physics, Augsburg
University, D-86135 Augsburg, Germany}

\author{Seunghyun Khim}
\affiliation{FPRD, Department of Physics and Astronomy, Seoul
National University, Seoul 151-742, Korea}

\author{J.~S.~Kim}
\affiliation{Departement of Physics, Pohang University of Science
and Technology, Pohang 790-784, Korea}

\author{Kee Hoon Kim}
\affiliation{FPRD, Department of Physics and Astronomy, Seoul
National University, Seoul 151-742, Korea}

\author{F.~Casper}
\affiliation{Institute for Inorganic and Analytic Chemistry,
Johannes Gutenberg-Universit\"{a}t, D-55099 Mainz, Germany}

\author{C.~Felser}
\affiliation{Institute for Inorganic and Analytic Chemistry,
Johannes Gutenberg-Universit\"{a}t, D-55099 Mainz, Germany}


\author{A.~Loidl}
\affiliation{Experimentalphysik V, Center for Electronic
Correlations and Magnetism, Institute for Physics, Augsburg
University, D-86135 Augsburg, Germany}

\date{\today}

\begin{abstract}
Electron spin resonance measurements in EuFe$_2$As$_2$ single
crystals revealed an absorption spectrum of a single resonance with
Dysonian lineshape. Above the spin-density wave transition at
$T_{\mathrm{SDW}}$ = 190~K the spectra are isotropic and the spin
relaxation is strongly coupled to the CEs resulting in a
Korringa-like increase of the linewidth. Below $T_{\mathrm{SDW}}$, a
distinct anisotropy develops and the relaxation behavior of the Eu
spins changes drastically into one with characteristic properties of
a magnetic insulating system, where dipolar and crystal-field
interactions dominate. This indicates a spatial confinement of the
CEs to the FeAs layers in the SDW state.

\end{abstract}


\pacs{76.30.-v,75.30.Fv,75.20.Hr,71.70.Ch}

\maketitle

The discovery of superconductivity in Fe-based pnictides and
chalcogenides has released an avalanche of scientific studies in
condensed-matter physics and chemistry. Three main material classes
are currently spurring the field: the \textit{R}FeAsO compounds with
\textit{R}=La-Gd (1111-systems) \cite{Kamihara08,Chen08}, the
ternary \textit{A}Fe$_2$As$_2$ class with \textit{A}=Ba,Sr,Ca,Eu
(122-systems) \cite{Rotter08a,Jeevan08b}, and the binary
chalcogenide systems such as FeSe
\cite{Hsu08,Mizuguchi08,Medvedev09}. The parent compounds of the
1111 and 122 systems exhibit a spin density wave (SDW) anomaly which
is accompanied by a structural distortion
\cite{Dong08,Rotter08b,Rotter09}. Upon doping the divalent
\textit{A}-site ions in the 122 compounds by monovalent ions like K
the SDW anomaly becomes suppressed and a superconducting ground
state appears.

Here we focus on EuFe$_2$As$_2$ which exhibits a SDW anomaly at
$T_{\rm{SDW}}$ = 190~K \cite{Raffius93,Jeevan08a,Wu09}. The
Eu$^{2+}$ ions with spin $S$ = 7/2 order antiferromagnetically at
$T_N$ = 19~K \cite{Ren08,Jeevan08a,Wu09}. The system reportedly
becomes superconducting upon substituting Eu by K \cite{Jeevan08b},
As by P \cite{Ren09}, or applying external pressure of about 26 kbar
\cite{Miclea09,Terashima09}.  In contrast to the other 122 systems,
where the substitution of Fe by Co also leads to superconductivity
\cite{Sefat08,Leithe-Jasper08}, the Eu compounds exhibit the onset
of a superconducting transition but seem to be hindered to reach
zero resistivity \cite{Zheng09}. It has been suggested that there is
a strong coupling between the localized Eu spins and the conduction
electrons (CEs) from the two-dimensional (2D) FeAs layers as
evidenced by magnetization and magneto-resistance measurements in
the parent compound \cite{Jiang09}.

To elucidate this coupling we investigated single crystalline
EuFe$_2$As$_2$ by electron spin resonance (ESR) spectroscopy. ESR
has been shown to be a highly sensitive tool to study the spin
fluctuations and magnetic interactions in cuprate superconductors
and their parent compounds (see e.g.~Ref.~\onlinecite{Elschner00}
and references therein). Our results show that above $T_{\rm{SDW}}$
the relaxation time of the Eu spins is dominated by the interaction
with the CEs, while the Eu system shows a relaxation behavior
reminiscent of a magnetic insulator below $T_{\rm{SDW}}$.


Polycrystalline EuFe$_2$As$_2$ was prepared following the procedure
described in \cite{Pfisterer83} and characterized by X-ray powder
diffraction using Mo-K$\alpha$ radiation ($\lambda$= 0.7093165~nm;
Bruker, AXS D8). High-quality EuFe$_2$As$_2$ single crystals were
grown using flux technique with starting composition of Eu:Fe:As:Sn
= 1:2:2:19, where Sn was removed by centrifugation after crystal
growth. The good quality of the single crystals was confirmed by
Laue x-ray diffraction as well as scanning electron microscopy
equipped with energy dispersive x-ray analysis. The in-plane
resistivity was measured using a standard 4-probe method. For
magnetization measurements we used a SQUID magnetometer MPMS5
(Quantum Design). ESR measurements were performed in a Bruker
ELEXSYS E500 CW-spectrometer at X-band frequencies ($\nu \approx$
9.36 GHz) equipped with a continuous He gas-flow cryostat in the
temperature region $4.2 < T< 300$~K. ESR detects the power $P$
absorbed by the sample from the transverse magnetic microwave field
as a function of the static magnetic field $H$. The signal-to-noise
ratio of the spectra is improved by recording the derivative $dP/dH$
using lock-in technique with field modulation.



\begin{figure}
\centering
\includegraphics[width=55mm,clip]{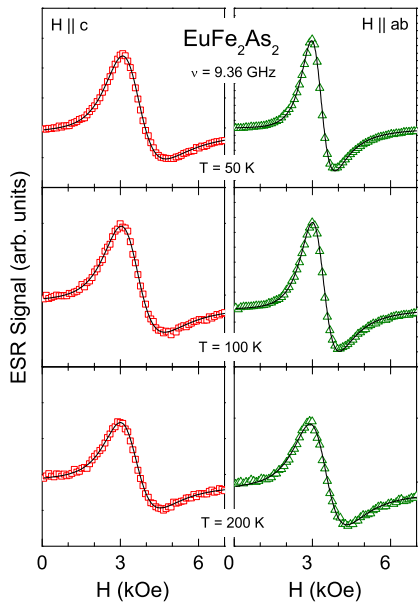}
\vspace{2mm} \caption[]{\label{spectra} (Color online) ESR spectra
of an EuFe$_2$As$_2$ single crystal taken at different temperatures
for the magnetic field applied parallel (left) and perpendicular
(right) to the $c$ axis.}
\end{figure}

Figure \ref{spectra} shows ESR spectra of EuFe$_2$As$_2$ for
different temperatures and orientations of the single crystal. In
all cases one observes a single exchange-narrowed resonance line
which is well described by a Dyson shape \cite{Barnes1981}
\begin{equation}
P(H) \propto \frac{\Delta H + \alpha (H-H_{\rm res})}{(H-H_{\rm
res})^2 + \Delta H^2},\label{Dyson}
\end{equation}
i.e. a Lorentz line at resonance field $H_{\rm res}$ with half width
at half maximum $\Delta H$ and a contribution $0 \leq \alpha \leq 1$
of dispersion to the absorption resulting in a characteristic
asymmetry. This is typical for metals where the skin effect drives
electric and magnetic components of the microwave field out of
phase. The dispersion to absorption (D/A) ratio $\alpha$ depends on
sample size, geometry, and skin depth. If the skin depth is small
compared to the sample size, $\alpha$ approaches 1. As $\Delta H$ is
of the same order of magnitude as $H_{\rm res}$, the counter
resonance at $-H_{\rm res}$ was included in the fitting process as
well \cite{JMR}.

\begin{figure}
\centering
\includegraphics[width=65mm,clip]{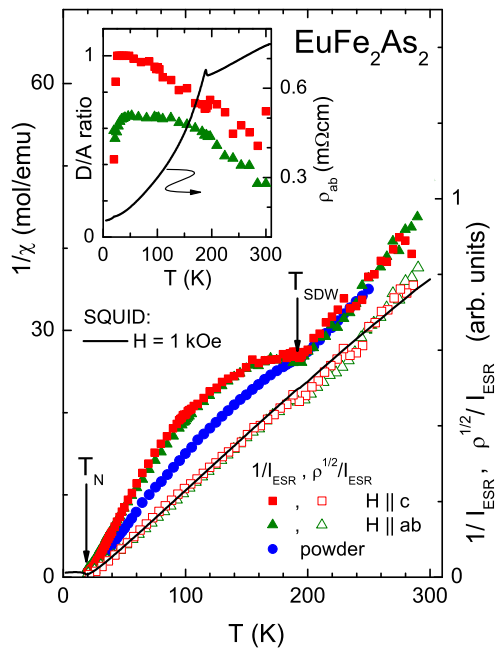}
\vspace{2mm} \caption[]{\label{daint} (Color online) $T$-dependence
of the inverse static susceptibility (left ordinate) of an
EuFe$_2$As$_2$ single crystal measured with $H=1000$\,Oe aligned in
the $ab$ plane, the inverse ESR intensity $1/I_{\rm ESR}$ (right
ordinate) for a single crystal and a powdered sample, and the
skin-depth corrected values for the single crystal data. Inset:
$T$-dependenc of the $D/A$-ratio and the in-plane resistivity of the
single crystal.}
\end{figure}

The $T$-dependence of the inverse double-integrated signal intensity
$I_{\rm ESR}$ is illustrated in Fig.~\ref{daint} for both prominent
orientations of the single crystal as well as for a powder sample.
Below 50~K and above 200~K, the inverse intensity 1/$I_{\rm ESR}$
follows a Curie-Weiss(CW)-like behavior with a CW temperature
$\Theta \approx 19$~K in agreement with the static susceptibility.
In the intermediate range one observes distinct deviations from
linearity, which is related to the changes of the D/A ratio and the
electrical resistivity shown in the inset of Fig.~\ref{daint}.  The
skin depth $\delta\propto \sqrt{\rho/\nu}$ and, hence, the partial
volume of the sample probed by the microwave field decreases with
decreasing temperature. These deviations are apparently reduced in
the powder sample, because the microwave field penetrates nearly all
sample volume due to the small grain size. After correction of the
single-crystal data with respect to the skin depth, we recover the
CW law in the entire temperature range (solid line in
Fig.~\ref{daint}).

\begin{figure}
\centering
\includegraphics[width=65mm,clip]{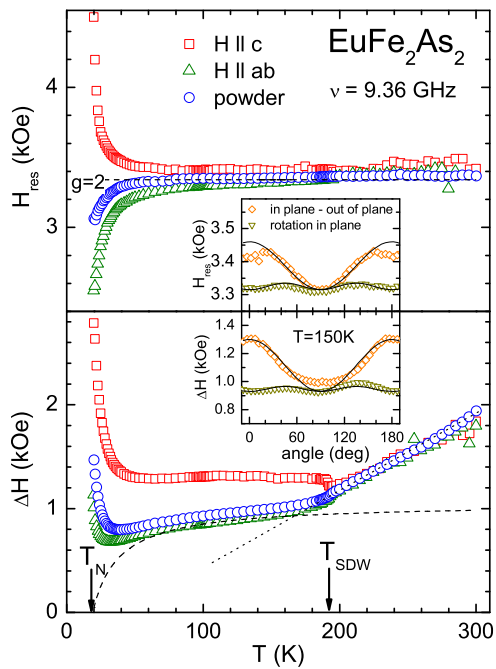}
\vspace{2mm} \caption[]{\label{dhres} (Color online) Temperature
dependence of the resonance field $H_{\rm res}$ (upper frame) and
linewidth $\Delta H$ (lower frame) of the ESR line in EuFe$_2$As$_2$
obtained for single crystal and powder sample. The insets illustrate
the anisotropies of $H_{\rm res}$ and $\Delta H$ for rotation of the
magnetic field within the $ab$ plane as well as from the $ab$ plane
to the $c$ direction. Solid lines $\propto
\cos^2\theta$,$\sin^2\theta$ are to guide the eyes, the dotted line
is linear fit, and the dashed line is a fit using Eq.~\ref{Huber}. }
\end{figure}

\begin{figure}
\centering
\includegraphics[width=60mm,clip]{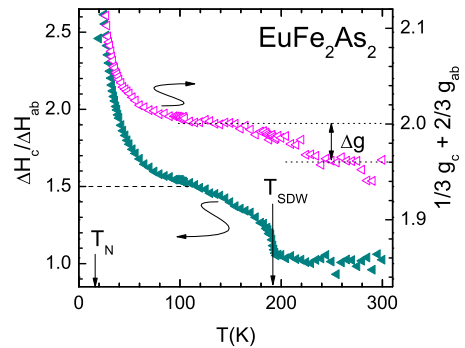}
\vspace{2mm} \caption[]{\label{dhr} (Color online) Temperature
dependence of the relative anisotropy of the resonance linewidth
$\Delta H_c/ \Delta H_{ab}$ (left ordinate, closed symbols) and the
averaged $g$ value (open symbols, right ordinate) of an
EuFe$_2$As$_2$ single crystal.}
\end{figure}

The $T$-dependences of $H_{\rm res}$ and $\Delta H$ are depicted in
Fig.~\ref{dhres} for $H\parallel c$ and $H\parallel ab$ and compared
to the corresponding data obtained in a powder sample. At high
temperature the ESR spectra are approximately isotropic at a
resonance field $H_{\rm res} \approx 3.41(3)$~kOe corresponding to a
$g$ value of 1.96(2), the linewidth increases linearly with
temperature with a slope of approximately 8~Oe/K. Below $T_{\rm
SDW}$ a pronounced anisotropy shows up in $H_{\rm res}$ and $\Delta
H$, which is illustrated in detail in the inset of Fig.~\ref{dhres}.
While a strong angular dependence with $180^{\circ}$ periodicity
appears, when rotating the field from the $c$ axis into the $ab$
plane, only a weak $90^{\circ}$ modulation is observed, when
rotating the field within the $ab$ plane. The former can be ascribed
to the dominant uniaxial crystal-electric field (CF) contribution,
which will be determined below. The latter indicates the
higher-order CF terms visible in the $ab$ plane, which will not be
further discussed here. On decreasing temperature the anisotropy
first tends to a kind of saturation (see Fig.~\ref{dhr}), but below
50~K it further diverges accompanied by a strong inhomogeneous
broadening towards $T_{\rm N}$ due to the onset of magnetic
fluctuations. In the following we will restrict the discussion to
temperatures $T>T_{\rm N}$.

%

\emph{Metallic regime for $T>T_{\rm{SDW}}$}: The ESR of local
moments in metals is characterized by a shift of the $g$ value
$\Delta g=J(0)N(E_{\rm F})$ from its value in insulators and a
linear increase of the linewidth $\Delta H \propto \langle J^2(q)
\rangle N^2(E_{\rm F}) T$ which both depend only on the
conduction-electron density of states $N(E_{\rm F})$ at the Fermi
energy $E_{\rm F}$ and the exchange constant $J$. The $g$ shift
results from the homogenous polarization of the CEs in the external
field (Pauli susceptibility), thus $J$ is taken at zero wave vector.
The linewidth is determined by the spin-flip scattering of CEs at
the local moments (Korringa relaxation) and, therefore, $J$ is
averaged over all possible scattering vectors $q$. Above $T_{\rm
SDW}$ the observed increase of the linewidth by 8~Oe/K is a typical
value for S-state $4f^7$ local moments in metals \cite{Taylor1975,
Barnes1981, Elschner1997} and, therefore, is ascribed to a pure
Korringa relaxation in a three-dimensional (3D) environment.

The negative $g$ shift $\Delta g \approx -0.04$ is unusual, but its
order of magnitude is typical for metals. The negative sign
indicates peculiarities of the 4\emph{f}-3\emph{d} coupling which
has been reported early in Gd doped Laves phases
\cite{Elschner1997}. It is remarkable that inspite of the tetragonal
symmetry of the crystal structure of EuFe$_2$As$_2$ resonance field
and linewidth are isotropic within experimental accuracy, i.e. the
CEs completely screen the ligand fields at the Eu site.

\emph{SDW state for $T<T_{\rm{SDW}}$}: Below $T_{\rm SDW}$ the
Korringa relaxation immediately disappears, although the resistivity
even decreases more strongly with decreasing temperature.
Concomitantly, the shift of the g value due to the polarization of
the CEs diminishes and the averaged \textit{g} value
$g=g_{c}/3+2g_{ab}/3$ below $T_{\rm SDW}$ corresponds to the typical
value $g=2.0$ for Eu$^{2+}$ in an insulating system. This is a first
indication that the formation of the SDW leads to a spatial
confinement of the CEs to the FeAs layers. Moreover, a pronounced
anisotropy shows up in the SDW state which reflects the symmetry of
the ligand fields. In this temperature regime, if not too close to
$T_{\rm N}$, the $T$-dependence of the linewidth can be well
described in terms of Eu spin-spin relaxation typical for magnetic
insulators. As pointed out by Huber \etal, in exchange coupled spin
systems the linewidth
\begin{equation}
\Delta H (T) = \frac{\chi_0}{\chi(T)} \Delta H_{\infty}
\label{Huber}
\end{equation}
is determined by the ratio of single-ion susceptibility $\chi_0
\propto 1/T$ and the experimental susceptibility $\chi(T)$ of
interacting spins multiplied by the high-temperature limit of the
linewidth $\Delta H_{\infty}$ \cite{Kubo1954,Huber1999}. This high
temperature limit can be estimated following the theory of exchange
narrowing of Anderson and Weiss \cite{Anderson53} as
\begin{equation}
\Delta H_{\infty} = \frac{h}{g \mu_{\rm B}} \frac{\langle \nu_{\rm
an}^2 \rangle}{\nu_{\rm ex}} \label{KT}
\end{equation}
where $\langle \nu_{\rm an}^2 \rangle$ denotes the second moment
of the resonance-frequency distribution due to any anisotropic
interaction like dipolar, hyperfine or crystal-electric field and
$\nu_{\rm ex}$ is the exchange frequency between the Eu spins.

The dipolar contribution to the second moment reads
\begin{equation}
\langle \nu^2_{\rm DD}\rangle = g^4 \mu_{\rm B}^4
\frac{3S(S+1)}{2h^2}\sum_{j\neq i} \frac{1+ \cos^2
\Theta_{ij}}{r_{ij}^6} \label{M2DDa}
\end{equation}
where $r_{ij}$ and $\Theta_{ij}$ denote the distance between spin
$i$ and $j$ and the polar angle of the external magnetic field with
respect to the direction of $r_{ij}$ \cite{Abragam1970}. The main
contribution results from the four nearest Eu neighbors at
$r_{ij}=a=3.907$\,\AA \cite{Jeevan08a}. With $g=2$ and $S=7/2$ one
obtains
\begin{equation}
\langle \nu^2_{\rm DD} \rangle (\Theta) = 36 {\rm \, GHz}^2 (2+
\sin^2 \Theta).\label{M2DDb}
\end{equation}
Here the polar angle $\Theta$ is measured between the direction of
the external field and the crystallographic $c$ axis. The exchange
constant $J$ between the Eu$^{2+}$ ions is determined from the CW
temperature $\Theta_{\rm CW} = 19$\,K using the Weiss molecular
field equations $3k_{\rm B} \Theta_{\rm CW} = J z S(S+1)$ with $z =
4$ exchange coupled nearest neighbors in the $ab$ plane as $J/k_{\rm
B} \approx 0.9$\,K. Then the exchange frequency can be approximately
estimated by $(h \nu_{\rm ex})^2 \approx z S(S+1)J^2$ resulting in
$\nu_{\rm ex} \approx 150$\,GHz. Thus the linewidth due to dipolar
broadening is determined as $\Delta H_{\infty} \approx 0.085 {\rm\,
kOe~}(2+\sin^2\Theta)$. This explains about 25\% of the experimental
linewidth, but exhibits an opposite anisotropy in comparison with
the experimental data. The hyperfine interaction in $^{151}$Eu
($^{151}A=103$\,MHz) and $^{153}$Eu ($^{153}A = 46$\,MHz)
\cite{Abragam1970} is at least one order of magnitude smaller than
the dipolar interaction, and hence can be neglected for the line
broadening. Therefore, only the tetragonal CF can account for the
observed anisotropy and magnitude of the linewidth.

The second moment of the leading uniaxial term of the CF is given by
\begin{equation}
\langle \nu^2_{\rm CF}\rangle (\Theta) = \frac{4S(S+1)-3}{10} D^2
(1 + \cos^2 \Theta)\label{M2CF}
\end{equation}
with the polar angle $\Theta$ measured between external field and
crystallographic $c$ axis \cite{Huber1999}. This provides the proper
anisotropy. The uniaxial zero-field splitting parameter $D$ can be
estimated from the experimentally observed asymptotic anisotropy of
the linewidth at intermediate temperatures (dashed line in
Fig.~\ref{dhr})
\begin{equation}
\frac{\Delta H_c}{\Delta H_{ab}} = \frac{\langle \nu^2_{\rm
CF}\rangle(0^{\circ}) + \langle \nu^2_{\rm
DD}\rangle(0^{\circ})}{\langle \nu^2_{\rm CF}\rangle (90^{\circ})
+ \langle \nu^2_{\rm DD}\rangle (90^{\circ})} \approx 1.5.
\end{equation}\label{aniso}
where $\langle \nu^2_{\rm an} \rangle = \langle \nu^2_{\rm DD}
\rangle + \langle \nu^2_{\rm CF} \rangle$ was assumed. Inserting
Eqs.~\ref{M2DDb} and \ref{M2CF} yields $D \approx 5.5$\,GHz, which
represents a reasonable order of magnitude \cite{KvN1998}. With the
obtained estimates for dipolar interaction, CF parameter and
exchange frequency one obtains $\Delta H_{\infty}(0^{\circ}) \approx
1.1$\,kOe in good agreement experiment.

In summary, the ESR properties in EuFe$_2$As$_2$ show distinct
differences between the high-temperature phase and the SDW state
below 190~K. Although the system remains metallic for all
temperatures, the ESR linewidth and $g$ value of the Eu$^{2+}$ ions
change from a typical behavior in a metallic environment with, e.g.,
a pure Korringa relaxation to characteristic anisotropic features as
usually observed in insulators (e.g.~spin-spin relaxation via
dipolar and crystal fields). We ascribe this abrupt change to a
local reduction of the 3D spin scattering due to a reduced
concentration of CEs at the Eu site and their spatial confinement to
the 2D FeAs layers in the SDW state.

 \emph{Note}. While finalizing this paper we became aware of an ESR
 study of Co doped EuFe$_2$As$_2$ for temperatures above 110~K, which
 shows a Korringa relaxation in agreement with our measurements \cite{Ying09}.

We thank Anna Pimenov for the SQUID measurements. We acknowledge
partial support
by the Deutsche Forschungsgemeinschaft (DFG) via the Collaborative
Research Center SFB 484 (Augsburg). The work at SNU was supported by
NRL (Grant No. M10600000238) program.



\end{document}